\documentstyle[epsf]{l-aa}
\def\co10{CO(1$\rightarrow$0)}
\def\co21{CO(2$\rightarrow$1)}
\newcount\fignumber\fignumber=1
\def\nfig{\global\advance\fignumber by 1}
\def\fignam#1{\xdef#1{\the\fignumber}}
\def\infig#1#2#3{\epsfxsize=#3cm \centering{\mbox{\epsfbox{#2}}}}
\fignam{\FWCOAGE}		\nfig		
\newcount\tabnumber\tabnumber=1
\def\ntab{\global\advance\tabnumber by 1}
\def\tabnam#1{\xdef#1{\the\tabnumber}}
\tabnam{\TWCOAGE}	\ntab		
\begin{document}
   \thesaurus{03(11.05.2, 11.09.4, 11.14.1, 11.19.3)}
\title{A correlation between CO linewidth and starburst age in barred spiral 
galaxies 
\thanks{Based on observations obtained at the 30 meter radiotelescope on Pico 
Veleta, operated by IRAM and the 1.93 meter telescope of Observatoire de
Haute-Provence, operated by INSU (CNRS).}
         }
   \author{T. Contini \inst{1} \and
           H. Wozniak \inst{2} \and
           S. Consid\`ere \inst{3} \and
           E. Davoust \inst{1}
           }
   \offprints{wozniak@observatoire.cnrs-mrs.fr}
   \institute{
Observatoire Midi-Pyr\'en\'ees, UMR 5572, 14 Avenue E. Belin, F-31400 Toulouse,
France 
\and 
Observatoire de Marseille, URA CNRS 237, 2 Place Le Verrier, F-13248 
Marseille Cedex 4, France
\and 
Observatoire de Besan\c{c}on, EP CNRS 123, B.P. 1615, F-25010 
Besan\c{c}on Cedex, France
             }
   \date{Received 15 october 1996; accepted 12 december 1996}
   \maketitle
   \begin{abstract} 
New CO(1$\rightarrow$0) and CO(2$\rightarrow$1) profiles complemented
by data from the literature are used to obtain CO linewidths for 29
barred spiral galaxies with young nuclear starbursts.  The ages of the
starbursts were estimated from optical spectroscopy and recent
evolutionary synthesis models.  The CO linewidths and the starburst
ages are correlated : galaxies with young (4-6 Myr) starbursts display
narrow ($\la 100$ km.s$^{-1}$) CO line while those with older
starbursts show broader CO lines. We discuss several scenarios of the
gas dynamics during the nuclear starbursts' evolution to interpret the
correlation.
	 \end{abstract}


\section{Introduction}
Our understanding of the links between the gas dynamics and the star
formation history in barred galaxies has made remarkable progress in
the past ten years, thanks to numerous multi-wavelength observations
and $N$-body simulations with stars, gas and star formation (Mihos \&
Hernquist 1994; Friedli \& Benz 1995).

For instance, the suggestion of Shlosman et al. (1989) that the gas
fuelling of the central regions is triggered by dissipation once the
bar is formed has been confirmed and the efficiency of the process has
been quantified (Friedli \& Benz 1993). However, in order to accrete
gas on a scale of a few parsec, another mechanism must be invoked,
like a secondary bar or a triaxial bulge (Shlosman et al. 1989,
1990). These secondary bars have now been observed in the optical
(Wozniak et al. 1995) and near infrared (Friedli et al. 1996) ranges,
and in the CO line (Kenney 1996 and references therein).  Friedli \&
Martinet (1993) have shown that double bar structures can transport
amounts of gas much closer to the galactic center than simple bars.
The gas accretion rate can be very high and may provide a possible
mechanism for triggering starbursts.

However, it still remains to be shown that AGN or nuclear starbursts
can be fuelled by such mechanisms. Indeed, high resolution
observations ({\it Hubble Space Telescope} images, CO mappings) are
beginning to reveal complex structures that simulations cannot
accurately model because of their poor resolution in the innermost
region.

In order to shed new light on the relationship between nuclear
starbursts and gas dynamics, we decided to combine the properties of
molecular clouds with characteristics of the starbursts rarely used in
previous studies. In particular, the age of the starburst has so far
never been used in practice; it can be now estimated from optical
spectroscopy and evolutionary synthesis models.

\section{Observations and data analysis}
\subsection{Sample selection and properties} 

We selected 29 galaxies (cf. Table~\TWCOAGE) from a large sample of
starburst galaxies (Contini 1996).  They share the following
properties: 1) they host a young nuclear starburst (age ranging from 3
to 12 Myr) and 2) they have the largest far infrared fluxes. The
latter criterion was to optimize our chances of detecting CO.

All galaxies of the sample are barred galaxies. This classification
comes from the LEDA database and/or from our CCD images at a
resolution of 1.5\arcsec\ (Contini 1996). Most galaxies have
morphological peculiarities, either an asymmetric spiral pattern (Mrk
133, 213, 353, 691, 799, 1466 and, to a lesser extent, Mrk 759 and
1050), several bright HII regions along the bar (Mrk 13, 281, 306,
710, 712, 731, 1341), or a polar ring (Mrk 306). Some galaxies also
have a close companion (Mrk 306, 602, 691, 1379). Mrk 2 has a
companion 103\arcsec\ away and Mrk 617 is a merger. The only galaxies
without marked peculiarities are Mrk 52, 412, 545, 575, 708, 1088,
1194 and 1485. Two galaxies lacked high resolution images (Mrk 1365
and 1379).

\subsection{CO line profiles}

The radio observations were obtained at the IRAM 30 meter
radiotelescope on August 20 to 24, 1995.  We observed in single
sideband at two frequencies simultaneously, CO(1$\rightarrow$0) at
115.3~Ghz with the 1.3mm SIS receiver, and CO(2$\rightarrow$1) at
230.5~Ghz with the 3mm (230G1) SIS receiver.  The beamwidth of the 30
meter antenna is 21\arcsec\ and 12\arcsec, and the main beam
efficiency 0.75 and 0.39 at 115 and 230~Ghz respectively.  The
observing procedure and data reduction are detailed in Contini (1996).

We observed 18 galaxies of the sample; CO data for the other 11
galaxies were taken from the literature. Mrk~712 is the only
undetected galaxy at 115~Ghz; this is not surprizing, as its far
infrared flux turns out to be the lowest of the whole sample. Mrk~412,
which also has a low far infrared flux, was barely detected at
115~GHz. Several galaxies, Mrk~13, 759, 1050, 1341, were ambiguously
detected at 230~Ghz, mainly because the noise at that frequency was a
factor 2 or 3 higher than at 115~GHz.  Mrk~213 was probably detected
at 230~Ghz, but the adopted recession velocity was too low, and,
consequently, part of the profile was outside the observing frequency
window.

The CO(1$\rightarrow$0) and CO(2$\rightarrow$1) line profiles are
displayed in Contini (1996). The spectra were smoothed to a final
velocity resolution of 10 km~s$^{-1}$ for both transitions. One of the
main characteristics of the profiles is that they are generally not
symmetric ; this cannot be attributed to noise, because of the long
integration times (typically between 50 and 330 minutes) and the good
resulting signal-to-noise ratio.  Most profiles can be adjusted by two
or three gaussians; this suggests the existence of separate molecular
clouds of uneven sizes and/or with distinct velocities. Only six
profiles (those of Mrk 13, 133, 412, 1050, 1379 and 1485) are well
fitted by one gaussian.

\begin{table} 
\caption[\TWCOAGE]{CO linewidths and 
starburst ages of the sample of barred Markarian galaxies. W(CO) is the 
width of the line estimated at 20 and 50\% of the maximum intensity of the 
CO profiles. Multiple entries correspond to different HII 
regions with comparable H$\beta$ fluxes} 
{\scriptsize
\begin{tabular}{rrrrrrrc}
\hline
\noalign{\smallskip}
Mrk&
\multicolumn{2}{c}{W(CO) 50\%}&
\multicolumn{2}{c}{W(CO) 20\%}&
\multicolumn{2}{c}{age}& Ref.\\
\noalign{\smallskip}
   &
(1$\rightarrow$0)&
(2$\rightarrow$1)&
(1$\rightarrow$0)&
(2$\rightarrow$1)&
   CMH94  &
  LH95&    \\ 
\noalign{\smallskip}
  &
\multicolumn{2}{c}{(km~s$^{-1}$)}&
\multicolumn{2}{c}{(km~s$^{-1}$)}&
\multicolumn{2}{c}{(Myr)}&  \\
\noalign{\smallskip}
\hline
\noalign{\smallskip}
   2& 106&  84& 133& 130& 5.6& 8.6&  \\
  13&  78&    & 122&    & 5.3& 4.8&  \\
 133&  83&  78&  94& 116& 4.5& 8.1&  \\
 213& 311&    & 389&    & 7.5& 7.2&  \\
 306&  67&  49& 161&  95& 7.4& 7.0&  \\
 353& 278& 259& 306& 330& 8.0& 9.0&  \\
 412&  78&    & 111&    & 5.6& 4.9&  \\
 575& 117& 108& 150& 149& 7.0& 8.5&  \\
 602& 194& 189& 278& 254& 5.7& 8.8&  \\
 691& 100&  59& 144& 138& 5.4& 4.9&  \\
 712&    &    &    &    &    &    &  \\
 731& 100&  57& 128&  73& 4.4& 7.9&  \\
 759& 167&    & 172&    & 9.0&11.0&  \\
1050& 250&    & 328&    & 8.0& 9.0&  \\
1341& 155&    & 200&    & 5.7& 8.7&  \\
1365& 217& 181& 267& 249& 9.8& 9.4&  \\
1379&  72&  59& 111&  95& 3.9& 4.5&  \\
    &    &    &    &    & 5.4& 4.9&  \\
    &    &    &    &    & 5.7& 8.6&  \\
1485& 240& 278& 361& 414&11.2&10.0&  \\
\hline
  52&  60&    & 133&    & 4.0& 6.3& 2 \\
 281& 278&    & 300&    & 6.5& 7.5& 1 \\
 534& 420&    & 482&    & 8.0& 8.9& 3 \\
 545& 364& 348& 473& 435& 6.0& 9.2& 5 \\
 617& 255& 287& 327& 348& 7.4& 8.5& 5 \\
 708& 187& 188& 275& 291& 5.7& 8.8& 4 \\
 710&    &  73&    & 137& 4.0& 6.3& 6 \\
    &    &    &    &    & 4.8& 7.7& 6 \\
    &    &    &    &    & 5.5& 8.4& 6 \\
 799& 309& 322& 400& 400& 6.0& 9.2&5,6\\
1088& 400& 157& 509& 365& 8.5& 9.2& 5 \\
1194& 291& 130& 364& 226& 9.0&11.0& 5 \\
1466& 127& 127& 182& 174& 5.3& 8.2& 5 \\
\noalign{\smallskip}
\hline
\multicolumn{8}{l}{\scriptsize References to CO linewidths (last col., bottom 
half of the Table):} \\ 
\multicolumn{8}{l}{\scriptsize 1 = Jackson et al. (1989); 2 = Young et al. 
(1995);} \\ 
\multicolumn{8}{l}{\scriptsize 3 = Wiklind \& Henkel (1989); 4 = Chini et al. 
(1992);} \\  
\multicolumn{8}{l}{\scriptsize 5 = Kr\"ugel et al. (1990); 6  = Contini et al. 
(1996).} \\   
\end{tabular}
}
\end{table}

The CO linewidths were measured directly on the profiles rather than
by a multi-gaussian fit, because, when several components are present
in the profile, the gaussians overlap and the FWHM thus does not
reflect the width of the profile. The linewidths were determined both
at 20 and 50\% of the maximum intensity for our sample as well as for
the data taken from the literature. We used these two estimates of the
linewidths to check the consistency of our results. For the galaxies
with high signal-to-noise ratio, the uncertainty on the linewidth is
equal to the velocity resolution (10~km~s$^{-1}$). The linewidths of
CO(1$\rightarrow$0) and CO(2$\rightarrow$1) profiles at 20 and 50\%
are given in Table~\TWCOAGE.  As the observing conditions (e.g. the
beamsize) and the calibration procedures for the 11 galaxies from the
literature were in general different from ours, these data are rather
heterogeneous and have been used with caution; they are thus listed
separately in Table~\TWCOAGE\ and plotted with different symbols in
Fig.~\FWCOAGE.

\subsection{Optical spectra and starburst ages}
The CCD spectra were obtained during several runs at the 1.93 meter
telescope of Observatoire de Haute-Provence, with the Carelec
spectrograph.  We took long-slit spectra at 260\AA/mm.  We observed
standard stars for flux calibration. The spectra were reduced with
MIDAS. The deblending of the H$\alpha$+[NII] and
[SII]$\lambda\lambda$6716,6731 lines was done by fitting multigaussian
profiles, with constraints on the relative positions of the individual
lines.

The starburst ages were estimated from the equivalent widths
W(H$\beta$) and metallicities (estimated from the oxygen abundance)
measured on our spectra and the evolutionary synthesis models of
Cervi\~no \& Mas-Hesse (1994, CMH94) and Leitherer \& Heckman (1995,
LH95). We used a standard Salpeter Initial Mass Function (IMF) with
$\alpha = 2.35$, $M_{\rm inf} = 1$ to 2 M$_\odot$ and $M_{\rm sup} =
100$ to 200 M$_\odot$. W(H$\beta$) was corrected for Balmer absorption
(assuming W$^{\rm abs}$(H$\beta$) $\simeq$ 2 \AA) which is a
contamination from the underlying stellar population.  The procedure
for estimating the starburst ages is fully described in Contini et
al. (1995) and Contini (1996). When several HII regions (or even a
fraction of a giant HII region) fall inside the beam of the
radiotelescope, the age is that of the brightest region in H$\alpha$.
HII regions outside the beam were not taken into
account. Incidentally, the brightest HII regions are all nuclear,
except for Mrk~13 (4.3\arcsec\ from the center) and Mrk~306
(7.8\arcsec\ from the center).

Three HII regions of Mrk~306 have similar (within a factor 2)
H$\alpha$ luminosities.  But the various ages are comparable for both
models (7.3 to 7.4 Myr using CMH94 and 7.0 to 8.5 Myr using LH95), so
that we adopted a mean value for the age. The three HII regions of
Mrk~710 also have similar H$\alpha$ luminosities, but the ages range
from 4 to 5.5~Myr (CMH94) and 6.3 to 8.4~Myr (LH95). We thus kept
three entries in our data for this object. For Mrk~1379, two HII
regions lie near the nucleus with very different ages (3.9, 5.7 using
CMH94; 4.5, 8.6 using LH95). Although the ratio of H$\alpha$
luminosities is close to 9, we decided to keep both regions in our
subsequent analysis.

A conservative estimate of the error on the age of young starbursts is
1~Myr for both models, taking into account the possible uncertainties
on the slope of the IMF, the contamination by older or younger HII
regions on the borderline of the beam, etc. For older starbursts,
H$\beta$ is weaker and the uncertainty on the age increases, and
reaches 2~Myr for 10~Myr old bursts.  However, there are several
galaxies for which the ages differ between CMH94 and LH95 models by
more than the sum of the age errors. {\it We thus decided to keep only
the objects showing a difference in age less than 2.5~Myr}. The final
age is thus the mean of the two estimates. This criterion does not
reduce the number of young starbursts displayed in Fig.~\FWCOAGE\ with
respect to those listed in Table~\TWCOAGE.  But fewer old starbursts
are displayed, since there the error on the age difference may reach
$\sim$4~Myr.

\begin{figure}
\infig{12}{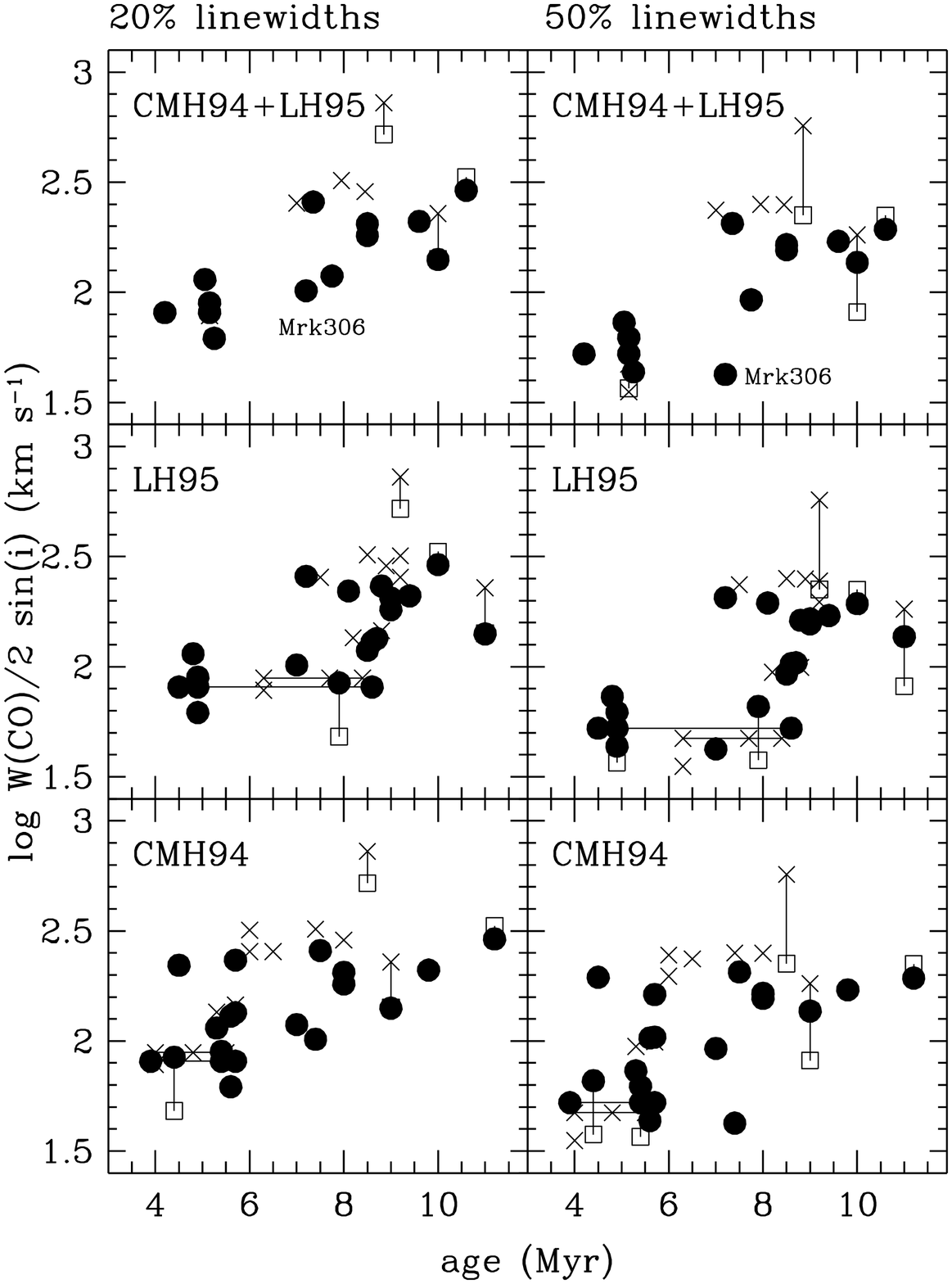}{8.8}
\caption[\FWCOAGE]{CO linewidth versus starburst age 
estimated from the models of LH95 ({\it middle} panel) and CMH94 ({\it bottom} 
panel).  The {\it top} panel is for the mean of the ages predicted by the two
models. $\bullet$: CO(1$\rightarrow$0) for our sample; $\times$:
CO(1$\rightarrow$0) for data from literature; $\Box$:
CO(2$\rightarrow$1), when it is different from CO(1$\rightarrow$0). 
Different data for the same object are joined by a line. 
Linewidths are measured at 20\% ({\it left} panels) and 50\% ({\it right} 
panels) of the maximum intensity of the CO profile
}
\end{figure}

\section{The CO linewidth -- starburst age correlation}
The CO linewidth as a function of starburst age is plotted in
Fig.~\FWCOAGE\ with the data of Table~\TWCOAGE.  When the two
linewidths are comparable, we only plot the one at 115~GHz, for the
sake of clarity. To correct for projection effects onto the plane of
the sky, we divided the observed width by $2\sin(i)$, where $i$ is the
inclination of the galaxy taken from the LEDA database.  For the top
diagram we used the {\it mean} of the two starburst ages determined in
the previous section.  We also show separate diagrams for the CMH94
and LH95 models, since they often give significantly different age
estimates.

A correlation appears in the diagrams using the mean and LH95
ages. Narrow CO linewidths (less than 100 km~s$^{-1}$) are associated
with young starbursts while older starbursts show greater linewidths
(between 150 and 300 km~s$^{-1}$).  There is one apparent exception,
Mrk 306, which has an older starburst and a narrow linewidth, but the
CO line profile shows several components, only one of which is taken
into account in the measure of the linewidth, because it is at least
twice as strong as the two others.

We have checked that the correlation between CO linewidth and
starburst age does not depend on the distance or on the absolute
magnitude (or, equivalently, the mass) of the galaxy; narrow
linewidths are not associated with low mass objects. We have also
verified that the CO linewidth is {\it independent} of the HI
linewidth.

The dispersion in Fig.~1 may have several causes. The dispersion in
starburst ages could be due to the contamination by younger
extra-nuclear HII regions.  It could also partly be due to errors on
the ages, which range from 1 to 2~Myr.  The dispersion in the CO
linewidths could be attributed to external causes like the interaction
with a companion. Let us recall that most galaxies of our sample are
peculiar in one way or another (cf. Sect.~2.1).  It could also be
partly due to the uncertainties on the angle $i$.

There could also be internal causes, such as the resolving power of
the radiotelescope at different frequencies. When the CO linewidth at
115 and at 230~Ghz are different, this may mean that the CO emission
is resolved at the lower frequency.  But different linewidth at the
two frequencies may also be due to different physical conditions; this
seems to be the case for Mrk 1485, for which the linewidth at the
higher frequency is larger than that at the lower frequency.

Furthermore, in Mrk~306, there is a great difference between the
CO(1$\rightarrow$0)\ and CO(2$\rightarrow$1)\ linewidths.  The
difference is due to the position of the most luminous HII region
which is 7.8\arcsec\ from the nucleus, inside the CO(1$\rightarrow$0)\
beam but outside the CO(2$\rightarrow$1)\ one; for this galaxy, we
only retained the CO(1$\rightarrow$0)\ data.

\section{Discussion}
Let us now interpret the correlation between CO linewidth and
starburst age. We restrict the discussion to the evolution of
starbursts and CO distributions in the central regions of galaxies,
since the starburst ages of our sample are those of {\it nuclear} HII
regions.

The first explanation that comes to mind is that the molecular clouds
are initially concentrated right near the nucleus ($r \leq$ 100
pc). The lines are broadened by galactic rotation, assuming a standard
rotation curve (Sofue 1996) and/or by the velocity dispersion among
individual clouds due to the kinetical energy from supernovae and
stellar winds associated with the starburst (Irwin \& Sofue 1996). As
the starburst ages, the gas expands outward ($r \sim$ 500 pc) and the
lines become wider because the clouds take part in the galactic
rotation. The problems with this scenario are that most of the gas
must stay in the disk and that there is no physical mechanism by which
the clouds can acquire enough angular momentum to participate to the
galactic rotation.  This is moreover in contradiction with the fact
that the gas on the contrary tends to lose angular momentum (in the
inflow scenario).

A classical gas inflow cannot explain the correlation either, since it
predicts that the starbursts may be ignited in a circum-nuclear ring
before the gas reaches the center, and young starbursts with broad CO
lines are expected.

Another possibility is that the clouds are pushed outward,
preferentially out of the plane of the galaxy, by powerful outflowing
winds generated by supernova explosions (Heckman et al. 1990).  The
broadening of the lines is then due to expansion rather than to
rotation.  The problem here is that the line profiles should be double
peaked, which is generally not the case in our sample.

Finally, the correlation could be spurious, if galaxies with young
starbursts are nearly face-on or have a low mass, but this is not the
case here.

None of the above scenarii, outflow of molecular gas inside or out of
the disk, are completely satisfactory.  This correlation remains a
puzzling problem which deserves more work. On one hand, the numerical
simulations must increase both their spatial and temporal resolutions
and include an energetic starburst in the nucleus of galaxies. On the
other hand, we must acquire higher resolution ($\la$ 100 pc) CO
observations in the nuclei of galaxies at
different stages of starburst evolution.

\begin{acknowledgements}
Data from the literature were obtained with the Lyon Meudon
Extragalactic database (LEDA), supplied by the LEDA team at
CRAL-Observatoire de Lyon (France).  We thank Rapha\"el Moreno (IRAM)
and the staff of OHP for assistance at the telescopes. We thank the referee,
I.~Shlosman, for fruitful comments.
\end{acknowledgements}

\end{document}